\documentclass[aps, 10pt,prl,twocolumn]{revtex4-1}
\usepackage{amssymb}
\usepackage{amsmath}
\usepackage{bm}
\usepackage{braket}
\usepackage[3D]{movie15}

\usepackage{graphicx}
 \usepackage{CJKutf8}
\usepackage[
pdfstartview=XYZ,
bookmarks=false,
colorlinks=true,
linkcolor=blue,
urlcolor=blue,
citecolor=blue,
pdftex,
bookmarks=false,
linktocpage=true, 
hyperindex=false
]{hyperref}

\usepackage{float}
\usepackage{mathtools}

\begin{document}
\title{Intrinsic two-dimensional states on the pristine surface of tellurium}
\author{Pengke Li (\begin{CJK*}{UTF8}{gbsn}李鹏科\end{CJK*})}
\affiliation{Department of Physics and Center for Nanophysics and Advanced Materials, U. Maryland, College Park, MD}
\author{Ian Appelbaum}
\affiliation{Department of Physics and Center for Nanophysics and Advanced Materials, U. Maryland, College Park, MD}

\begin{abstract}
Atomic chains configured in a helical geometry have fascinating properties, including phases hosting localized bound states in their electronic structure. We show how the zero-dimensional state -- bound to the edge of a single one-dimensional helical chain of tellurium atoms -- evolves into two-dimensional bands on the $c$-axis surface of the three-dimensional trigonal bulk. We give an effective Hamiltonian description of its dispersion in $k$-space by exploiting confinement to a virtual bilayer, and elaborate on the diminished role of spin-orbit coupling. These previously-unidentified intrinsic gap-penetrating surface bands were neglected in the interpretation of seminal experiments, where two-dimensional transport was otherwise attributed to extrinsic accumulation layers.     
\end{abstract}
\maketitle

{\em{Introduction--}} The bulk lattice of the elemental chalcogenide tellurium (Te) is formed from van~der~Waals-bonded 3-fold helices aligned to the [0001] direction ($c$-axis) of the trigonal crystal \cite{Doi_JPSJ70,Du_NanoLett17}. Interhelix coupling and translational symmetry cause the eigenstates in each one-dimensional helix to disperse as a function of Bloch wavevector in the plane perpendicular to the $c$-axis, forming the bulk band structure of the three-dimensional solid. Many intriguing features are found in this spectrum, such as a valence-band spin texture enabling circular photogalvanic effect and current-induced longitudinal spin polarization along the $c$-axis \cite{Ivchenko_JETPL78}, and the ability to form topological insulator \cite{Agapito_PRL13} or Weyl nodes \cite{Hirayama_PRL15} under strain sufficient to close the bulk band gap.

\begin{figure}[h!]

\def\fn{Te_cface}
\raggedright
\vspace{0.25in}
\mbox{\hspace{0.1in}}
\includemovie[
	poster,
	toolbar, 
	label=\fn.u3d,
	text={},
3Droo=30,
3Dcoo=0.9183776378631592 -0.0006829077610746026 -3.589715003967285,
3Dc2c=-0.6930592656135559 -0.4559812545776367 0.5583457350730896,
3Droll=1.7430079271266015,
3Dlights=CAD,
3Daac=30
]{1.25in}{1.25in}{\fn.u3d}%

\raisebox{1.55in}[3.25in][3in]{\includegraphics[width=3.25in]{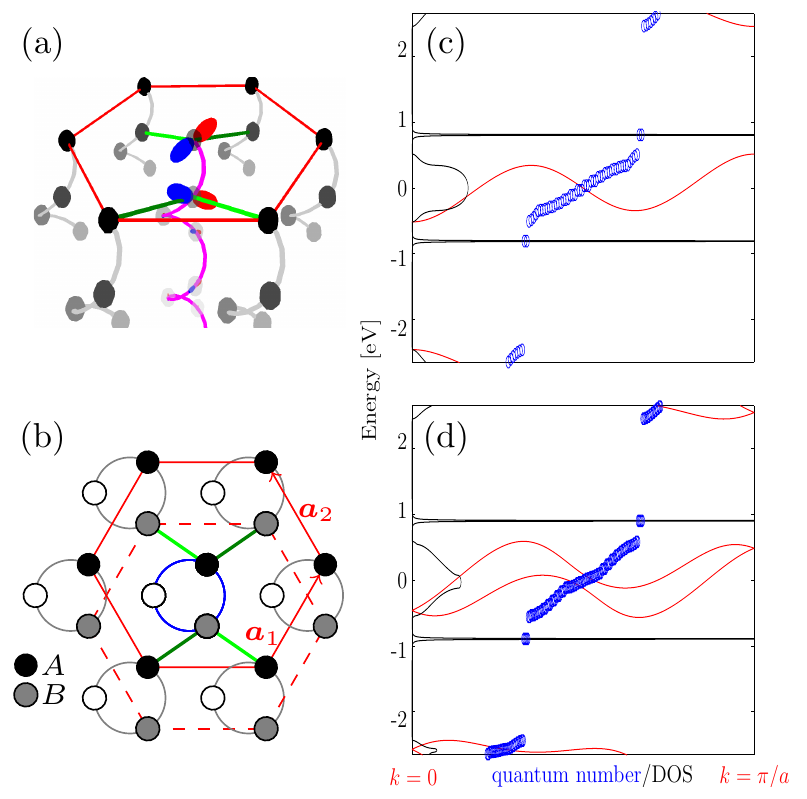}}
\vspace{-4.5in}

\caption{(a) [Interactive 3D Online] Schematic view of $c$-axis surface of the trigonal tellurium lattice formed by van~der~Waals bonded 3-fold helices. A red hexagon connects the six surface atoms, highlighted by darker shading, surrounding a central helix in purple. Orbitals represent the local amplitude and phase for the gapped bound-state wavefunction of a single helix that disperses into a surface band upon coupling with nearest neighbors via light and dark green interhelix bonds. (b) Plan view, showing three atoms in the surface unit cell. Solid/dashed hexagon highlights surface/sub-surface sites denoted as type $A/B$ on nearest-neighbor helices. Contribution from the third site (open circle) in the unit cell is diminished by negligible wavefunction amplitude.  (c)/(d) Energy spectrum for a single 1-dimensional helical lattice of tellurium atoms, without/with on-site spin-orbit coupling. Bloch bands in red, discrete spectrum of 50-site finite lattice in blue, and local density of states at the endpoint of a semi-infinite lattice in black. The wavefunction orbital components of the gapped bound state in (c) at $\approx |V_{\pi}^{(1)}|=0.8$~eV is superimposed on the central helix in (a). \label{fig:1}
}
\end{figure}

When each helix is finite, as shown in Fig. \ref{fig:1}(a), a $c$-axis surface is present at the termination site. If localized edge modes are bound to the terminated endpoints, these too will couple and disperse, forming two-dimensional surface states \cite{Tamm_PZSU32,Shockley_PR39}. In previous work, we identified strict conditions for a topological phase hosting zero-energy  edge-bound modes in a generic single atomic helix \cite{Li_PRB17}. In addition to that robust zero-energy mode (related to the Su-Schrieffer-Heeger topological state in dimerized chains \cite{Su_PRB80}) embedded within a continuum band, a pair of gapped states above and below the band appear in an appropriate parameter range.

In the present paper, we show how these zero-dimensional gapped edge modes associated with the trigonal helix in tellurium disperse into two-dimensional surface states that penetrate deep into the band gap from the valence band. Their presence has important experimental consequences \cite{Laude_PRL72}, overlooked in the interpretation of  results from several classic papers on tellurium \cite{vonKlitzing_SSC71,Tsiulyanu_B11,Silbermann_JJAP74,Englert_PSS77} that played a seminal role in the development of two-dimensional transport physics \cite{vonKlitzing_ARCMP17}.

{\em{A single helix--}} In Ref.~[\onlinecite{Li_PRB17}], we investigated the spectrum of single helical atomic chains with a general rotational symmetry and pitch. Our analytic approach relied upon strict chiral symmetry, where on-site terms in the Hamiltonian, such as local disorder and spin-orbit interaction (SOI), are absent. In the current study, we immediately recognize that the high atomic number of tellurium demands that these more realistic factors must be carefully taken into account.

Our approach starts by comparing the bandstructure of an infinite helical chain of tellurium atoms calculated without and with SOI, as shown by the red curves in Fig.~\ref{fig:1}(c) and (d), respectively. In the nearest-neighbor tight-binding formalism, the three orthogonal $p$-orbitals are used as basis functions of the Slater-Koster Hamiltonian \cite{Slater_PR54} with two coupling parameters $V_{\sigma}^{(1)}=2.9$~eV and $V_{\pi}^{(1)}=-0.8$~eV. The spin-orbit coupling (SOC) manifest in Fig.~\ref{fig:1}(d) is induced by on-site atomic SOI with $\lambda=0.4$~eV \cite{Chadi_PRB77}.

As a group-VI elemental semiconductor, Te has six outer shell electrons filling two $s$ states  and four $p$ states. Taking spin degeneracy into account, the Fermi energy lies in the gap above the central bulk band in the three-band $p$-orbital model shown in Fig.~\ref{fig:1}(c). 
The consequence of terminating the lattice at an edge is captured by the density of states (DOS) spectrum for a semi-infinite helix evaluated from the surface Green's function \cite{Dy_PRB79, Appelbaum_PRB04}, shown in black. 
Chiral symmetry of the Hamiltonian results in the apparent even symmetry in energy. 
Most importantly, there are two endpoint-localized bound states in the bulk gap, indicated by the broadened $\delta$-functions at energy $\approx\pm V_{\pi}^{(1)}$, and explicitly verified by the discrete spectrum of a finite 50-site helical chain shown in blue. 
These gapped states are entirely intrinsic, and in no way due to extrinsic factors such as band-bending resulting from charge depletion.
They are therefore expected to persist in the presence of perturbative factors such as the inclusion of remote orbitals in an expanded basis \cite{Joannopoulos_PRB75}, finite disorder in the bulk, and edge decorations (irregular hopping of the edge bond or adatoms of other types), unless these perturbations push the edge state energies into the bulk spectrum.

The orbital components of the wavefunction for the spin-independent bound state at energy close to $+|V_{\pi}^{(1)}|$ is depicted in Fig.~\ref{fig:1}(a), superimposed on the central helix lattice. It is clear that most of the electron probability lies at the first two atoms close to the surface, labeled $A$ and $B$ in Fig.~\ref{fig:1}(b). This segregation can be understood by recognizing that, in addition to the decay of any bound state wavefunction from the edge into the bulk, states close to $\pm V_{\pi}^{(1)}$ (an eigenvalue of the hopping matrix) have nearly `tripartite' spatial distribution as a generic result of the nearest neighbor tight binding model \cite{Li_PRB17},  suppressing the wavefunction amplitude at the third site. This fact will be used to construct a compact effective Hamiltonian for the [0001] surface state on bulk tellurium formed from these zero-dimensional bound modes. 

The three-site periodicity of the generic tripartite state with energy $\pm|V_\pi^{(1)}|$ is key to explaining the emergence of a bound state from the delocalized spectrum of one-dimensional states in the first place. Consider the case in which the band edge lies at this energy, with no gapped bound state. Band extrema under this condition must be doubly degenerate with $k=\pm\frac{\pi}{3a}$, allowing construction of a $\phi(z_n)=\sin (\frac{\pi}{3a}z_n)$ wavefunction envelope. When these bulk states can be induced to recede from $\pm|V_\pi^{(1)}|$ by tuning the Hamiltonian parameters, a bound state with this form [acquiring an exponential decay but still satisfying the $\phi(z_n = 0)=0$ boundary condition at the edge] is exposed in the gap.

\begin{figure}
\includegraphics[width=3.25in]{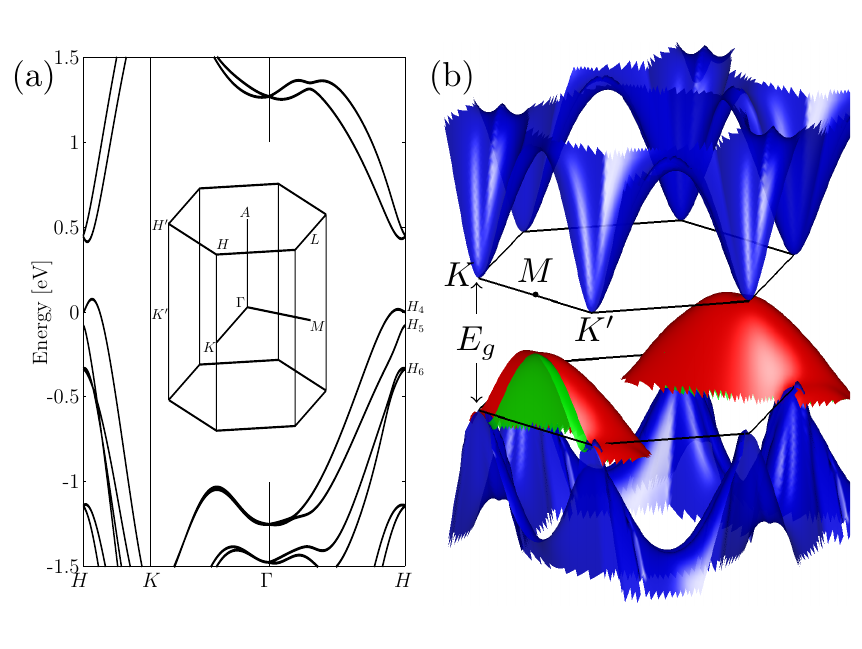}
\caption{Electronic band structure of tellurium using a full $p_{x,y,z}$-orbital tight-binding basis. Panel (a) shows bulk bands, (b) shows the projection of a slab spectrum onto the two-dimensional c-axis surface Brillouin zone. Blue surfaces indicate valence and conduction bands delocalized in the bulk; red and green surfaces penetrating the gap highlight the two spin-split bands confined to a single surface. 
\label{fig:BZ}}
\end{figure}

The robustness of these gapped boundary states is evident by their preservation even after SOI is included in Fig. \ref{fig:1}(d), where SOC is manifest in the bulk bands by clear lifting of spin degeneracy at all $k$-points not possessing time-reversal invariance. Furthermore, energetic symmetry in the DOS is lost due to broken chiral Hamiltonian symmetry. In contrast, the two edge states are only slightly shifted in energy by SOI. 

The weakness of SOC in the edge states is also reflected in their wavefunctions. In comparison with the bulk states, SOI induces a much smaller spin mixing, which can be understood from a perturbation point of view. Starting from the spinless edge state wavefunction, the inclusion of atomic on-site SOI as a perturbation involves matrix elements with states at remote energies. Although they have high spatial overlap, the pair of edge states at $\approx \pm V_\pi^{(1)}$ consist of exactly the same amplitudes of orbital components (differing only by opposite phase on every other site due to the chiral symmetry of the unperturbed spinless Hamiltonian), so that their mutual on-site SOI matrix element vanishes. Nonzero contributions to SOC in the surface state are therefore induced through all other intermediate states at remote energies \textit{only} within the bulk spectrum, whose spatial wavefunction overlap with the edge state is quite limited. We will readdress this SOC weakness later when discussing the [0001] surface state on bulk tellurium, into which these bound states evolve.

{\em{Bulk electronic structure--}} The real Te crystal is constructed by aligning identical helical chains in a trigonal Bravais lattice with lattice constant $a = 4.44$~\AA. In addition to the tight binding Hamiltonian of a single helix, the coupling between neighboring atomic chains is captured by  second-nearest-neighbor hopping between $p$-orbitals of sites on adjacent chains as shown by the green interhelix bonds in Fig. \ref{fig:1}(a) and (b), with Slater-Koster coupling parameters $V^{(2)}_{\sigma}= 0.7$~eV and $V^{(2)}_{\pi} = 0.2$~eV \cite{Joannopoulos_PRB75}. Salient features evident in the bulk bandstructure using these parameters [see Fig. \ref{fig:BZ}(a)] are the $E_g\approx0.335$~eV bandgap, frontier states close to the $H$-point, $H_4-H_5-H_6$ valence band splitting (around 0.1~eV and 0.2~eV), and appropriate effective masses at the relevant band extrema.

{\em{Surface electronic structure--}} 
Threefold rotations in the bulk crystal space group require partial translation along the helix. On the $c$-axis surface, however, only in-plane symmetry operations preserve the lattice, while partial translation along the helix and the 2-fold rotation of the $D_3$ point group that maintains invariance of the bulk are forbidden. As a result, other than Kramers' degeneracy, surface band energies in the reduced 2D hexagonal Brillouin zone (BZ) are no longer degenerate at otherwise equivalent $\bm{k}$-points of high symmetry, so evaluation of states throughout the full hexagonal surface BZ is necessary.

In Fig.~\ref{fig:BZ}(b), 
we present the result from a tight-binding electronic structure calculation of a 60-atom-thick $c$-axis-oriented slab, where the surface bands of only one of the two (top or bottom) surfaces are isolated and appear in red and green. These spin-split surface bands are anisotropic domes centered around one of the three $M$ points, determined by the termination phase within the surface unit cell, while the blue surfaces depict the bulk valence and conduction bands. These surface bands arise from the transverse dispersion of the single helix edge state close to the valence band and extend far into the band gap (by $\approx 0.2$ eV) along one particular $K-M-K^\prime$ path. 

\begin{figure}
\includegraphics[width=3.25in]{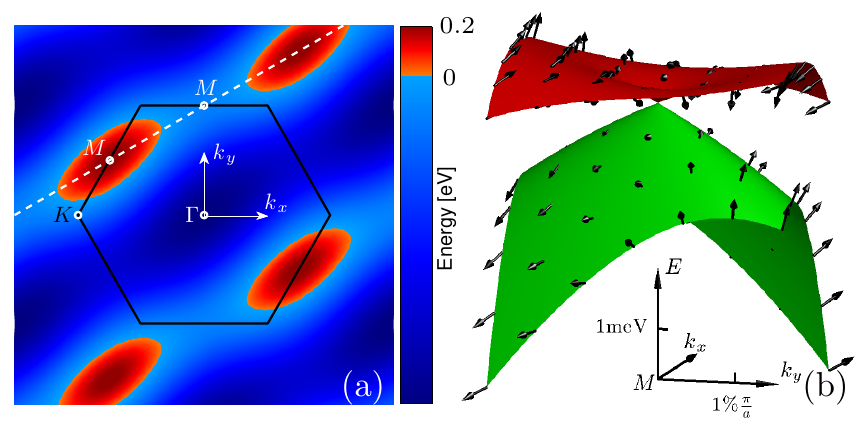}%
\caption{(a) Analytic model of surface state dispersion, Eq.~\ref{eq:analytic}, with matrix elements calculated from the bound state of a single helix without spin-orbit interaction; see Fig. \ref{fig:1}(a). Red indicates states above the valence band maximum, penetrating into the bulk gap.  Panel (b) shows anisotropic spin orbit splitting close to the $M$-point degeneracy. Colors here correspond to assignments in Fig. \ref{fig:BZ}(b). The directions and lengths of the arrows indicate the orientations of the spin-eigenstates, and the strengths of the spin-orbit field, respectively.
\label{fig:analytic}}
\end{figure}

The wavefunctions of these two-dimensional states are localized to the slab surface, and exponentially decay into the bulk. This behavior is reminiscent of their origin in the zero-dimensional bound states of a single helix, whose wavefunction is dominated by orbitals localized at the A and B sites closest to the edge. Therefore, we can consider an approximation consisting only of a bilayer trigonal 2D lattice as shown by the solid and dashed red hexagons in Fig.~\ref{fig:1}(b), occupied by $\ket{A}$ and $\ket{B}$ orbitals shown in Fig. \ref{fig:1}(a), respectively. Using this simplification, a lowest-order spinless model for the 2D surface band can be constructed directly from A/B-site interchain coupling. The dispersion is given by
\begin{align}
E_{\text{surf}}(\bm{k})=2U_1\cos(\bm{k}\cdot \bm{a}_1)+2U_2\cos(\bm{k}\cdot \bm{a}_2),
\label{eq:analytic}
\end{align}
where $\bm{a}_{1,2}$ are lattice vectors given in Fig.~\ref{fig:1}(b). $U_{1} = -0.3$~eV and $U_{2}=-0.09$~eV are Slater-Koster matrix elements of the orbital components of the single helix gapped edge state, and characterize the interhelix  coupling strength\cite{Slater_PR54} along the light and dark green bonds, shown in Fig. \ref{fig:1}(b). As plotted throughout the Brillouin zone in Fig. \ref{fig:analytic}(a), Eq.~(\ref{eq:analytic}) clearly captures the most salient features of the surface band in the full slab calculation [Fig. \ref{fig:BZ}(b)] such as the gap penetration and mass anisotropy.

An expansion of Eq.~(\ref{eq:analytic}) can be used to estimate the anisotropic effective mass around the extrema at the $M$-point $\bm{k}_M=[-\frac{\pi}{a},\frac{\pi}{\sqrt{3}a}]$, where the dispersion approximates an elliptic paraboloid. Since the first term with $U_1$ coefficient  dominates in Eq.~(\ref{eq:analytic}), the major axis is nearly perpendicular to the lattice vector $\bm{a}_1$, along the line connecting this $M$-point to the adjacent one at $[0,\frac{2\pi}{\sqrt{3}a}]$ ($30^\circ$ from the $k_x$ axis). To lowest order in $U_2/U_1$, the anisotropic effective masses are then
\begin{align}
m^*_{||}=\frac{2\hbar^2}{3a^2U_2}\approx 2.9m_0,\quad 
m^*_{\perp}=\frac{2\hbar^2}{a^2(4U_1+U_2)}\approx 0.6m_0,
\notag
\end{align}
giving a DOS effective mass of $m^*=\sqrt{m^*_{||}m^*_{\perp}}\approx 1.3m_0$, and a two-dimensional DOS of $\frac{m^*}{\pi\hbar^2}\approx 5.5\times 10^{14}$cm$^{-2}/$eV. We stress that these values are meaningful only in the vicinity of the dispersion maxima well into the bulk band gap, and may not reflect the empirically measured dynamical mass values if the chemical potential lies close to the bulk valence band, where nonparabolicities are significant.

As in the case of a single helix, SOI plays a significant role in the electronic structure of bulk Te, reflected by the large split-off energies [comparable with the band gap, see Fig.~\ref{fig:BZ}(a)] between the \{$H_4$, $H_5$, $H_6$\} valence bands originating from the orbitally degenerate $H_3$ representation in single-group \cite{Doi_JPSJ70}. However, at the crystal boundary, SOC is much less pronounced, as evident in the relatively weak spin splitting of surface bands shown in Fig. \ref{fig:BZ}(b). This again indicates the robustness of the gapped edge states originating from the coupling of neighboring sites in the bulk of a single helix, and the weakness of SOC at the edge can likewise be understood via second order perturbation theory. 

We thus include SOC in our analytic model of the surface band effective Hamiltonian by first numerically calculating the single-helix spin-dependent edge states. These already capture spin-mixing from the delocalized states, enabling us to derive the surface state dispersion in the transverse Brillouin zone by Slater-Koster interhelix coupling. To lowest order, the general form of SOC close to the surface band maximum is anisotropically linear in the wavevector and takes the general form $k_i v_{ij}\sigma_j$, where $i=\{x,y\}$ but $j = \{x,y,z\}$. Here, $\sigma_j$ are the $2\times 2$ Pauli matrices, and the $2\times 3$ tensor $v_{ij}$ represents the strength of the spin-orbit field as shown in Fig.~\ref{fig:analytic}(b). All tensor components differ but are of the same meV$\cdot$\AA\, scale.

{\em{Conclusion--}} We have shown that the dispersion of the intrinsic surface state on the $c$-axis of tellurium is determined almost exclusively by the wavefunction of a bound mode at the last two atomic sites of a helical chain. The surface state can thus effectively be thought of as a group-VI elemental virtual bilayer two-dimensional electronic material, in addition to group-IV graphene, silicene \cite{LYV_PRB07}, germanene \cite{Takeda_PRB94}, and group-V phosphorene \cite{Li_PRB14}. However, because the two atoms are strongly coupled to the bulk through intra-helix bonds, there is no flexural phonon that typically plagues the charge and spin scattering mobilities in exfoliated layers \cite{Song_PRL13}. Nevertheless, the transport limitations of the Te surface state remain to be investigated \cite{Averkiev_FTT96, *Averkiev_PSS96, Averkiev_FTT98, *Averkiev_PSS98}.

Just above Te in the chalcogen group, selenium (Se) also forms in the trigonal crystalline phase. As shown in Supplementary Materials, similar tight-binding calculations show that $c$-face surface bands, analogous to those investigated here, are present in this material as well, despite different lattice constants and Slater-Koster parameters needed to reproduce the larger bulk gap and frontier states near the $A$-point.    

The surface state we describe requires a pristine surface exactly perpendicular to the $c$-axis. Since its properties are dependent on coupling between the first two atomic sites from the surface, we expect that atomic-scale termination disorder or subtle surface reconstruction will drastically affect its presence. Even steps on otherwise perfect vicinal surfaces will tend to decouple conducting states on domains of locally perfect termination. These concerns impose constraints on experiments designed to test our theoretical predictions. Probes of electronic structure \cite{Nakayama_PRB17} with high spatial resolution are necessary, such as with STM quasiparticle interference \cite{Crommie_Nature93,Huang_NanoLett17}. Unfortunately, thin films of Te typically grow with the $c$-axis in the substrate plane \cite{Weidmann_TSF71}, so bulk sample preparation may be required. 

Finally, we draw attention to work by von~Klitzing \& Landwehr on Shubnikov-de Haas measurements (magnetic field-induced quantum oscillations in resistivity) associated with a two-dimensional hole gas on the trigonal Te $c$-face surface \cite{vonKlitzing_SSC71}. These observations were attributed to subbands within an accumulation layer on the c-axis surface, created electrostatically via band-bending \cite{Huang_NanoLett17}, nominally plausible due to ``valence alteration pair'' defects \cite{Tsiulyanu_B11}. Although two Fermi surfaces were identified for measurements on the bare surface, field-effect gating sometimes produced four \cite{Silbermann_JJAP74}. In later work, it was shown that gate electric fields are incapable of restoring flat-band conditions necessary to eliminate the accumulation layer, despite field strengths $\mathcal{E}_{max}$ in the range of MV/cm \cite{Englert_PSS77}. Only after  chemical treatment of the surface increased the sample resistance by an order of magnitude was observation of a transconductance minimum obtained. 

The present theoretical description of an intrinsic surface state on the pristine $c$-face clearly does not originate from any extrinsic effects that would otherwise be necessary to support the original explanation for such experimental observations. The large DOS and gap penetration implies a total charge accommodated in states above the bulk valence band maximum of $\approx 10^{14}$cm$^{-2}$, explaining the inability of field-effect to appreciably tune the chemical potential (i.e. $\epsilon \mathcal{E}_{max}\approx 10^{12}$cm$^{-2}$).   We are thus compelled to suggest that the interpretation of these decades-old pioneering experimental measurements  is revisited. 

\begin{acknowledgments}
We thank J. Sau for several enlightening discussions. We acknowledge support from the Office of Naval Research under contract N000141712994, and the Defense Threat Reduction Agency under contract HDTRA1-13-1-0013. 
\end{acknowledgments}


\begin{thebibliography}{32}%
\makeatletter
\providecommand \@ifxundefined [1]{%
 \@ifx{#1\undefined}
}%
\providecommand \@ifnum [1]{%
 \ifnum #1\expandafter \@firstoftwo
 \else \expandafter \@secondoftwo
 \fi
}%
\providecommand \@ifx [1]{%
 \ifx #1\expandafter \@firstoftwo
 \else \expandafter \@secondoftwo
 \fi
}%
\providecommand \natexlab [1]{#1}%
\providecommand \enquote  [1]{``#1''}%
\providecommand \bibnamefont  [1]{#1}%
\providecommand \bibfnamefont [1]{#1}%
\providecommand \citenamefont [1]{#1}%
\providecommand \href@noop [0]{\@secondoftwo}%
\providecommand \href [0]{\begingroup \@sanitize@url \@href}%
\providecommand \@href[1]{\@@startlink{#1}\@@href}%
\providecommand \@@href[1]{\endgroup#1\@@endlink}%
\providecommand \@sanitize@url [0]{\catcode `\\12\catcode `\$12\catcode
  `\&12\catcode `\#12\catcode `\^12\catcode `\_12\catcode `\%12\relax}%
\providecommand \@@startlink[1]{}%
\providecommand \@@endlink[0]{}%
\providecommand \url  [0]{\begingroup\@sanitize@url \@url }%
\providecommand \@url [1]{\endgroup\@href {#1}{\urlprefix }}%
\providecommand \urlprefix  [0]{URL }%
\providecommand \Eprint [0]{\href }%
\providecommand \doibase [0]{http://dx.doi.org/}%
\providecommand \selectlanguage [0]{\@gobble}%
\providecommand \bibinfo  [0]{\@secondoftwo}%
\providecommand \bibfield  [0]{\@secondoftwo}%
\providecommand \translation [1]{[#1]}%
\providecommand \BibitemOpen [0]{}%
\providecommand \bibitemStop [0]{}%
\providecommand \bibitemNoStop [0]{.\EOS\space}%
\providecommand \EOS [0]{\spacefactor3000\relax}%
\providecommand \BibitemShut  [1]{\csname bibitem#1\endcsname}%
\let\auto@bib@innerbib\@empty
\bibitem [{\citenamefont {Doi}\ \emph {et~al.}(1970)\citenamefont {Doi},
  \citenamefont {Nakao},\ and\ \citenamefont {Kamimura}}]{Doi_JPSJ70}%
  \BibitemOpen
  \bibfield  {author} {\bibinfo {author} {\bibfnamefont {T.}~\bibnamefont
  {Doi}}, \bibinfo {author} {\bibfnamefont {K.}~\bibnamefont {Nakao}}, \ and\
  \bibinfo {author} {\bibfnamefont {H.}~\bibnamefont {Kamimura}},\ }\href
  {\doibase 10.1143/JPSJ.28.36} {\bibfield  {journal} {\bibinfo  {journal} {J.
  Phys. Soc. Jpn.}\ }\textbf {\bibinfo {volume} {28}},\ \bibinfo {pages} {36}
  (\bibinfo {year} {1970})}\BibitemShut {NoStop}%
\bibitem [{\citenamefont {Du}\ \emph {et~al.}(2017)\citenamefont {Du} \emph
  {et~al.}}]{Du_NanoLett17}%
  \BibitemOpen
  \bibfield  {author} {\bibinfo {author} {\bibfnamefont {Y.}~\bibnamefont {Du}}
  \emph {et~al.},\ }\href {\doibase 10.1021/acs.nanolett.7b01717} {\bibfield
  {journal} {\bibinfo  {journal} {Nano Lett.}\ }\textbf {\bibinfo {volume}
  {17}},\ \bibinfo {pages} {3965} (\bibinfo {year} {2017})}\BibitemShut
  {NoStop}%
\bibitem [{\citenamefont {Ivchenko}\ and\ \citenamefont
  {Pikus}(1978)}]{Ivchenko_JETPL78}%
  \BibitemOpen
  \bibfield  {author} {\bibinfo {author} {\bibfnamefont {E.}~\bibnamefont
  {Ivchenko}}\ and\ \bibinfo {author} {\bibfnamefont {G.}~\bibnamefont
  {Pikus}},\ }\href {http://www.jetpletters.ac.ru/ps/1554/article_23792.shtml}
  {\bibfield  {journal} {\bibinfo  {journal} {JETP Lett.}\ }\textbf {\bibinfo
  {volume} {27}},\ \bibinfo {pages} {604} (\bibinfo {year} {1978})}\BibitemShut
  {NoStop}%
\bibitem [{\citenamefont {Agapito}\ \emph {et~al.}(2013)\citenamefont
  {Agapito}, \citenamefont {Kioussis}, \citenamefont {Goddard},\ and\
  \citenamefont {Ong}}]{Agapito_PRL13}%
  \BibitemOpen
  \bibfield  {author} {\bibinfo {author} {\bibfnamefont {L.~A.}\ \bibnamefont
  {Agapito}}, \bibinfo {author} {\bibfnamefont {N.}~\bibnamefont {Kioussis}},
  \bibinfo {author} {\bibfnamefont {W.~A.}\ \bibnamefont {Goddard}}, \ and\
  \bibinfo {author} {\bibfnamefont {N.~P.}\ \bibnamefont {Ong}},\ }\href
  {\doibase 10.1103/PhysRevLett.110.176401} {\bibfield  {journal} {\bibinfo
  {journal} {Phys. Rev. Lett.}\ }\textbf {\bibinfo {volume} {110}},\ \bibinfo
  {pages} {176401} (\bibinfo {year} {2013})}\BibitemShut {NoStop}%
\bibitem [{\citenamefont {Hirayama}\ \emph {et~al.}(2015)\citenamefont
  {Hirayama}, \citenamefont {Okugawa}, \citenamefont {Ishibashi}, \citenamefont
  {Murakami},\ and\ \citenamefont {Miyake}}]{Hirayama_PRL15}%
  \BibitemOpen
  \bibfield  {author} {\bibinfo {author} {\bibfnamefont {M.}~\bibnamefont
  {Hirayama}}, \bibinfo {author} {\bibfnamefont {R.}~\bibnamefont {Okugawa}},
  \bibinfo {author} {\bibfnamefont {S.}~\bibnamefont {Ishibashi}}, \bibinfo
  {author} {\bibfnamefont {S.}~\bibnamefont {Murakami}}, \ and\ \bibinfo
  {author} {\bibfnamefont {T.}~\bibnamefont {Miyake}},\ }\href {\doibase
  10.1103/PhysRevLett.114.206401} {\bibfield  {journal} {\bibinfo  {journal}
  {Phys. Rev. Lett.}\ }\textbf {\bibinfo {volume} {114}},\ \bibinfo {pages}
  {206401} (\bibinfo {year} {2015})}\BibitemShut {NoStop}%
\bibitem [{\citenamefont {Tamm}(1932)}]{Tamm_PZSU32}%
  \BibitemOpen
  \bibfield  {author} {\bibinfo {author} {\bibfnamefont {I.}~\bibnamefont
  {Tamm}},\ }\href@noop {} {\bibfield  {journal} {\bibinfo  {journal} {Phys. Z.
  Soviet Union}\ }\textbf {\bibinfo {volume} {1}},\ \bibinfo {pages} {733}
  (\bibinfo {year} {1932})}\BibitemShut {NoStop}%
\bibitem [{\citenamefont {Shockley}(1939)}]{Shockley_PR39}%
  \BibitemOpen
  \bibfield  {author} {\bibinfo {author} {\bibfnamefont {W.}~\bibnamefont
  {Shockley}},\ }\href {\doibase 10.1103/PhysRev.56.317} {\bibfield  {journal}
  {\bibinfo  {journal} {Phys. Rev.}\ }\textbf {\bibinfo {volume} {56}},\
  \bibinfo {pages} {317} (\bibinfo {year} {1939})}\BibitemShut {NoStop}%
\bibitem [{\citenamefont {Li}\ \emph {et~al.}(2017)\citenamefont {Li},
  \citenamefont {Sau},\ and\ \citenamefont {Appelbaum}}]{Li_PRB17}%
  \BibitemOpen
  \bibfield  {author} {\bibinfo {author} {\bibfnamefont {P.}~\bibnamefont
  {Li}}, \bibinfo {author} {\bibfnamefont {J.~D.}\ \bibnamefont {Sau}}, \ and\
  \bibinfo {author} {\bibfnamefont {I.}~\bibnamefont {Appelbaum}},\ }\href
  {\doibase 10.1103/PhysRevB.96.115446} {\bibfield  {journal} {\bibinfo
  {journal} {Phys. Rev. B}\ }\textbf {\bibinfo {volume} {96}},\ \bibinfo
  {pages} {115446} (\bibinfo {year} {2017})}\BibitemShut {NoStop}%
\bibitem [{\citenamefont {Su}\ \emph {et~al.}(1980)\citenamefont {Su},
  \citenamefont {Schrieffer},\ and\ \citenamefont {Heeger}}]{Su_PRB80}%
  \BibitemOpen
  \bibfield  {author} {\bibinfo {author} {\bibfnamefont {W.~P.}\ \bibnamefont
  {Su}}, \bibinfo {author} {\bibfnamefont {J.~R.}\ \bibnamefont {Schrieffer}},
  \ and\ \bibinfo {author} {\bibfnamefont {A.~J.}\ \bibnamefont {Heeger}},\
  }\href {\doibase 10.1103/PhysRevB.22.2099} {\bibfield  {journal} {\bibinfo
  {journal} {Phys. Rev. B}\ }\textbf {\bibinfo {volume} {22}},\ \bibinfo
  {pages} {2099} (\bibinfo {year} {1980})}\BibitemShut {NoStop}%
\bibitem [{\citenamefont {Laude}\ \emph {et~al.}(1972)\citenamefont {Laude},
  \citenamefont {Willis},\ and\ \citenamefont {Fitton}}]{Laude_PRL72}%
  \BibitemOpen
  \bibfield  {author} {\bibinfo {author} {\bibfnamefont {L.~D.}\ \bibnamefont
  {Laude}}, \bibinfo {author} {\bibfnamefont {R.~F.}\ \bibnamefont {Willis}}, \
  and\ \bibinfo {author} {\bibfnamefont {B.}~\bibnamefont {Fitton}},\ }\href
  {\doibase 10.1103/PhysRevLett.29.472} {\bibfield  {journal} {\bibinfo
  {journal} {Phys. Rev. Lett.}\ }\textbf {\bibinfo {volume} {29}},\ \bibinfo
  {pages} {472} (\bibinfo {year} {1972})}\BibitemShut {NoStop}%
\bibitem [{\citenamefont {von Klitzing}\ and\ \citenamefont
  {Landwehr}(1971)}]{vonKlitzing_SSC71}%
  \BibitemOpen
  \bibfield  {author} {\bibinfo {author} {\bibfnamefont {K.}~\bibnamefont {von
  Klitzing}}\ and\ \bibinfo {author} {\bibfnamefont {G.}~\bibnamefont
  {Landwehr}},\ }\href {\doibase 10.1016/0038-1098(71)90630-2} {\bibfield
  {journal} {\bibinfo  {journal} {Solid State Commun.}\ }\textbf {\bibinfo
  {volume} {9}},\ \bibinfo {pages} {2201 } (\bibinfo {year}
  {1971})}\BibitemShut {NoStop}%
\bibitem [{\citenamefont {Tsiulyanu}(2011)}]{Tsiulyanu_B11}%
  \BibitemOpen
  \bibfield  {author} {\bibinfo {author} {\bibfnamefont {D.}~\bibnamefont
  {Tsiulyanu}},\ }\enquote {\bibinfo {title} {Tellurium thin films in sensor
  technology},}\ in\ \href {\doibase 10.1007/978-94-007-0903-4_38} {\emph
  {\bibinfo {booktitle} {Nanotechnological Basis for Advanced Sensors}}},\
  \bibinfo {editor} {edited by\ \bibinfo {editor} {\bibfnamefont {J.~P.}\
  \bibnamefont {Reithmaier}}, \bibinfo {editor} {\bibfnamefont
  {P.}~\bibnamefont {Paunovic}}, \bibinfo {editor} {\bibfnamefont
  {W.}~\bibnamefont {Kulisch}}, \bibinfo {editor} {\bibfnamefont
  {C.}~\bibnamefont {Popov}}, \ and\ \bibinfo {editor} {\bibfnamefont
  {P.}~\bibnamefont {Petkov}}}\ (\bibinfo  {publisher} {Springer Netherlands},\
  \bibinfo {address} {Dordrecht},\ \bibinfo {year} {2011})\ pp.\ \bibinfo
  {pages} {363--380}\BibitemShut {NoStop}%
\bibitem [{\citenamefont {Silbermann}\ \emph {et~al.}(1974)\citenamefont
  {Silbermann}, \citenamefont {Landwehr}, \citenamefont {Thuillier},\ and\
  \citenamefont {Bouat}}]{Silbermann_JJAP74}%
  \BibitemOpen
  \bibfield  {author} {\bibinfo {author} {\bibfnamefont {R.}~\bibnamefont
  {Silbermann}}, \bibinfo {author} {\bibfnamefont {G.}~\bibnamefont
  {Landwehr}}, \bibinfo {author} {\bibfnamefont {J.~C.}\ \bibnamefont
  {Thuillier}}, \ and\ \bibinfo {author} {\bibfnamefont {J.}~\bibnamefont
  {Bouat}},\ }\href {http://stacks.iop.org/1347-4065/13/i=S2/a=359} {\bibfield
  {journal} {\bibinfo  {journal} {Jpn. J. Appl. Phys.}\ }\textbf {\bibinfo
  {volume} {13}},\ \bibinfo {pages} {359} (\bibinfo {year} {1974})}\BibitemShut
  {NoStop}%
\bibitem [{\citenamefont {Englert}\ \emph {et~al.}(1977)\citenamefont
  {Englert}, \citenamefont {von Klitzing}, \citenamefont {Silbermann},\ and\
  \citenamefont {Landwehr}}]{Englert_PSS77}%
  \BibitemOpen
  \bibfield  {author} {\bibinfo {author} {\bibfnamefont {T.}~\bibnamefont
  {Englert}}, \bibinfo {author} {\bibfnamefont {K.}~\bibnamefont {von
  Klitzing}}, \bibinfo {author} {\bibfnamefont {R.}~\bibnamefont {Silbermann}},
  \ and\ \bibinfo {author} {\bibfnamefont {G.}~\bibnamefont {Landwehr}},\
  }\href {\doibase 10.1002/pssb.2220810110} {\bibfield  {journal} {\bibinfo
  {journal} {Phys. Stat. Solidi (b)}\ }\textbf {\bibinfo {volume} {81}},\
  \bibinfo {pages} {119} (\bibinfo {year} {1977})}\BibitemShut {NoStop}%
\bibitem [{\citenamefont {von Klitzing}(2017)}]{vonKlitzing_ARCMP17}%
  \BibitemOpen
  \bibfield  {author} {\bibinfo {author} {\bibfnamefont {K.}~\bibnamefont {von
  Klitzing}},\ }\href {\doibase 10.1146/annurev-conmatphys-031016-025148}
  {\bibfield  {journal} {\bibinfo  {journal} {Annu. Rev. Condens. Matter
  Phys.}\ }\textbf {\bibinfo {volume} {8}},\ \bibinfo {pages} {13} (\bibinfo
  {year} {2017})}\BibitemShut {NoStop}%
\bibitem [{\citenamefont {Slater}\ and\ \citenamefont
  {Koster}(1954)}]{Slater_PR54}%
  \BibitemOpen
  \bibfield  {author} {\bibinfo {author} {\bibfnamefont {J.~C.}\ \bibnamefont
  {Slater}}\ and\ \bibinfo {author} {\bibfnamefont {G.~F.}\ \bibnamefont
  {Koster}},\ }\href {\doibase 10.1103/PhysRev.94.1498} {\bibfield  {journal}
  {\bibinfo  {journal} {Phys. Rev.}\ }\textbf {\bibinfo {volume} {94}},\
  \bibinfo {pages} {1498} (\bibinfo {year} {1954})}\BibitemShut {NoStop}%
\bibitem [{\citenamefont {Chadi}(1977)}]{Chadi_PRB77}%
  \BibitemOpen
  \bibfield  {author} {\bibinfo {author} {\bibfnamefont {D.~J.}\ \bibnamefont
  {Chadi}},\ }\href {\doibase 10.1103/PhysRevB.16.790} {\bibfield  {journal}
  {\bibinfo  {journal} {Phys. Rev. B}\ }\textbf {\bibinfo {volume} {16}},\
  \bibinfo {pages} {790} (\bibinfo {year} {1977})}\BibitemShut {NoStop}%
\bibitem [{\citenamefont {Dy}\ \emph {et~al.}(1979)\citenamefont {Dy},
  \citenamefont {Wu},\ and\ \citenamefont {Spratlin}}]{Dy_PRB79}%
  \BibitemOpen
  \bibfield  {author} {\bibinfo {author} {\bibfnamefont {K.~S.}\ \bibnamefont
  {Dy}}, \bibinfo {author} {\bibfnamefont {S.-Y.}\ \bibnamefont {Wu}}, \ and\
  \bibinfo {author} {\bibfnamefont {T.}~\bibnamefont {Spratlin}},\ }\href
  {\doibase 10.1103/PhysRevB.20.4237} {\bibfield  {journal} {\bibinfo
  {journal} {Phys. Rev. B}\ }\textbf {\bibinfo {volume} {20}},\ \bibinfo
  {pages} {4237} (\bibinfo {year} {1979})}\BibitemShut {NoStop}%
\bibitem [{\citenamefont {Appelbaum}\ \emph {et~al.}(2004)\citenamefont
  {Appelbaum}, \citenamefont {Wang}, \citenamefont {Joannopoulos},\ and\
  \citenamefont {Narayanamurti}}]{Appelbaum_PRB04}%
  \BibitemOpen
  \bibfield  {author} {\bibinfo {author} {\bibfnamefont {I.}~\bibnamefont
  {Appelbaum}}, \bibinfo {author} {\bibfnamefont {T.}~\bibnamefont {Wang}},
  \bibinfo {author} {\bibfnamefont {J.~D.}\ \bibnamefont {Joannopoulos}}, \
  and\ \bibinfo {author} {\bibfnamefont {V.}~\bibnamefont {Narayanamurti}},\
  }\href {\doibase 10.1103/PhysRevB.69.165301} {\bibfield  {journal} {\bibinfo
  {journal} {Phys. Rev. B}\ }\textbf {\bibinfo {volume} {69}},\ \bibinfo
  {pages} {165301} (\bibinfo {year} {2004})}\BibitemShut {NoStop}%
\bibitem [{\citenamefont {Joannopoulos}\ \emph {et~al.}(1975)\citenamefont
  {Joannopoulos}, \citenamefont {Schl\"uter},\ and\ \citenamefont
  {Cohen}}]{Joannopoulos_PRB75}%
  \BibitemOpen
  \bibfield  {author} {\bibinfo {author} {\bibfnamefont {J.~D.}\ \bibnamefont
  {Joannopoulos}}, \bibinfo {author} {\bibfnamefont {M.}~\bibnamefont
  {Schl\"uter}}, \ and\ \bibinfo {author} {\bibfnamefont {M.~L.}\ \bibnamefont
  {Cohen}},\ }\href {\doibase 10.1103/PhysRevB.11.2186} {\bibfield  {journal}
  {\bibinfo  {journal} {Phys. Rev. B}\ }\textbf {\bibinfo {volume} {11}},\
  \bibinfo {pages} {2186} (\bibinfo {year} {1975})}\BibitemShut {NoStop}%
\bibitem [{\citenamefont {Guzm\'an-Verri}\ and\ \citenamefont {Lew
  Yan~Voon}(2007)}]{LYV_PRB07}%
  \BibitemOpen
  \bibfield  {author} {\bibinfo {author} {\bibfnamefont {G.~G.}\ \bibnamefont
  {Guzm\'an-Verri}}\ and\ \bibinfo {author} {\bibfnamefont {L.~C.}\
  \bibnamefont {Lew Yan~Voon}},\ }\href {\doibase 10.1103/PhysRevB.76.075131}
  {\bibfield  {journal} {\bibinfo  {journal} {Phys. Rev. B}\ }\textbf {\bibinfo
  {volume} {76}},\ \bibinfo {pages} {075131} (\bibinfo {year}
  {2007})}\BibitemShut {NoStop}%
\bibitem [{\citenamefont {Takeda}\ and\ \citenamefont
  {Shiraishi}(1994)}]{Takeda_PRB94}%
  \BibitemOpen
  \bibfield  {author} {\bibinfo {author} {\bibfnamefont {K.}~\bibnamefont
  {Takeda}}\ and\ \bibinfo {author} {\bibfnamefont {K.}~\bibnamefont
  {Shiraishi}},\ }\href {\doibase 10.1103/PhysRevB.50.14916} {\bibfield
  {journal} {\bibinfo  {journal} {Phys. Rev. B}\ }\textbf {\bibinfo {volume}
  {50}},\ \bibinfo {pages} {14916} (\bibinfo {year} {1994})}\BibitemShut
  {NoStop}%
\bibitem [{\citenamefont {Li}\ and\ \citenamefont
  {Appelbaum}(2014)}]{Li_PRB14}%
  \BibitemOpen
  \bibfield  {author} {\bibinfo {author} {\bibfnamefont {P.}~\bibnamefont
  {Li}}\ and\ \bibinfo {author} {\bibfnamefont {I.}~\bibnamefont {Appelbaum}},\
  }\href {\doibase 10.1103/PhysRevB.90.115439} {\bibfield  {journal} {\bibinfo
  {journal} {Phys. Rev. B}\ }\textbf {\bibinfo {volume} {90}},\ \bibinfo
  {pages} {115439} (\bibinfo {year} {2014})}\BibitemShut {NoStop}%
\bibitem [{\citenamefont {Song}\ and\ \citenamefont {Dery}(2013)}]{Song_PRL13}%
  \BibitemOpen
  \bibfield  {author} {\bibinfo {author} {\bibfnamefont {Y.}~\bibnamefont
  {Song}}\ and\ \bibinfo {author} {\bibfnamefont {H.}~\bibnamefont {Dery}},\
  }\href {\doibase 10.1103/PhysRevLett.111.026601} {\bibfield  {journal}
  {\bibinfo  {journal} {Phys. Rev. Lett.}\ }\textbf {\bibinfo {volume} {111}},\
  \bibinfo {pages} {026601} (\bibinfo {year} {2013})}\BibitemShut {NoStop}%
\bibitem [{\citenamefont {Averkiev}\ and\ \citenamefont
  {Pikus}(1996{\natexlab{a}})}]{Averkiev_FTT96}%
  \BibitemOpen
  \bibfield  {author} {\bibinfo {author} {\bibfnamefont {N.}~\bibnamefont
  {Averkiev}}\ and\ \bibinfo {author} {\bibfnamefont {G.}~\bibnamefont
  {Pikus}},\ }\href@noop {} {\bibfield  {journal} {\bibinfo  {journal} {Fiz.
  Tverd. Tela}\ }\textbf {\bibinfo {volume} {38}},\ \bibinfo {pages} {1748}
  (\bibinfo {year} {1996}{\natexlab{a}})}\BibitemShut {NoStop}%
\bibitem [{\citenamefont {Averkiev}\ and\ \citenamefont
  {Pikus}(1996{\natexlab{b}})}]{Averkiev_PSS96}%
  \BibitemOpen
  \bibfield  {author} {\bibinfo {author} {\bibfnamefont {N.}~\bibnamefont
  {Averkiev}}\ and\ \bibinfo {author} {\bibfnamefont {G.}~\bibnamefont
  {Pikus}},\ }\href@noop {} {\bibfield  {journal} {\bibinfo  {journal} {Phys.
  Solid State}\ }\textbf {\bibinfo {volume} {38}},\ \bibinfo {pages} {964}
  (\bibinfo {year} {1996}{\natexlab{b}})}\BibitemShut {NoStop}%
\bibitem [{\citenamefont {Averkiev}\ and\ \citenamefont
  {Pikus}(1998{\natexlab{a}})}]{Averkiev_FTT98}%
  \BibitemOpen
  \bibfield  {author} {\bibinfo {author} {\bibfnamefont {N.}~\bibnamefont
  {Averkiev}}\ and\ \bibinfo {author} {\bibfnamefont {G.}~\bibnamefont
  {Pikus}},\ }\href@noop {} {\bibfield  {journal} {\bibinfo  {journal} {Fiz.
  Tverd. Tela}\ }\textbf {\bibinfo {volume} {39}},\ \bibinfo {pages} {1659}
  (\bibinfo {year} {1998}{\natexlab{a}})}\BibitemShut {NoStop}%
\bibitem [{\citenamefont {Averkiev}\ and\ \citenamefont
  {Pikus}(1998{\natexlab{b}})}]{Averkiev_PSS98}%
  \BibitemOpen
  \bibfield  {author} {\bibinfo {author} {\bibfnamefont {N.}~\bibnamefont
  {Averkiev}}\ and\ \bibinfo {author} {\bibfnamefont {G.}~\bibnamefont
  {Pikus}},\ }\href@noop {} {\bibfield  {journal} {\bibinfo  {journal} {Phys.
  Solid State}\ }\textbf {\bibinfo {volume} {39}},\ \bibinfo {pages} {1481}
  (\bibinfo {year} {1998}{\natexlab{b}})}\BibitemShut {NoStop}%
\bibitem [{\citenamefont {Nakayama}\ \emph {et~al.}(2017)\citenamefont
  {Nakayama} \emph {et~al.}}]{Nakayama_PRB17}%
  \BibitemOpen
  \bibfield  {author} {\bibinfo {author} {\bibfnamefont {K.}~\bibnamefont
  {Nakayama}} \emph {et~al.},\ }\href {\doibase 10.1103/PhysRevB.95.125204}
  {\bibfield  {journal} {\bibinfo  {journal} {Phys. Rev. B}\ }\textbf {\bibinfo
  {volume} {95}},\ \bibinfo {pages} {125204} (\bibinfo {year}
  {2017})}\BibitemShut {NoStop}%
\bibitem [{\citenamefont {Crommie}\ \emph {et~al.}(1993)\citenamefont
  {Crommie}, \citenamefont {Lutz},\ and\ \citenamefont
  {Eigler}}]{Crommie_Nature93}%
  \BibitemOpen
  \bibfield  {author} {\bibinfo {author} {\bibfnamefont {M.}~\bibnamefont
  {Crommie}}, \bibinfo {author} {\bibfnamefont {C.}~\bibnamefont {Lutz}}, \
  and\ \bibinfo {author} {\bibfnamefont {D.~M.}\ \bibnamefont {Eigler}},\
  }\href {\doibase 10.1038/363524a0} {\bibfield  {journal} {\bibinfo  {journal}
  {Nature}\ }\textbf {\bibinfo {volume} {363}},\ \bibinfo {pages} {524 }
  (\bibinfo {year} {1993})}\BibitemShut {NoStop}%
\bibitem [{\citenamefont {Huang}\ \emph {et~al.}(2017)\citenamefont {Huang}
  \emph {et~al.}}]{Huang_NanoLett17}%
  \BibitemOpen
  \bibfield  {author} {\bibinfo {author} {\bibfnamefont {X.}~\bibnamefont
  {Huang}} \emph {et~al.},\ }\href {\doibase 10.1021/acs.nanolett.7b01029}
  {\bibfield  {journal} {\bibinfo  {journal} {Nano Lett.}\ }\textbf {\bibinfo
  {volume} {17}},\ \bibinfo {pages} {4619} (\bibinfo {year}
  {2017})}\BibitemShut {NoStop}%
\bibitem [{\citenamefont {Weidmann}\ and\ \citenamefont
  {Anderson}(1971)}]{Weidmann_TSF71}%
  \BibitemOpen
  \bibfield  {author} {\bibinfo {author} {\bibfnamefont {E.}~\bibnamefont
  {Weidmann}}\ and\ \bibinfo {author} {\bibfnamefont {J.}~\bibnamefont
  {Anderson}},\ }\href
  {http://www.sciencedirect.com/science/article/pii/0040609071900733?via%3Dihub}
  {\bibfield  {journal} {\bibinfo  {journal} {Thin\,\,{Solid}\,\,Films}\
  }\textbf {\bibinfo {volume} {7}},\ \bibinfo {pages} {265} (\bibinfo {year}
  {1971})}\BibitemShut {NoStop}%
\end{thebibliography}

%

\end{document}